\pdfoutput=1 
\documentclass[conference]{sheffcontrol}

\usepackage{cite}
\usepackage[tbtags]{amsmath}
\usepackage{amssymb,amsfonts}
\usepackage{algorithmic}
\usepackage{graphicx}
\usepackage{textcomp}
\usepackage[utf8]{inputenc}
\usepackage{tikz}
\usetikzlibrary{shapes,arrows}
\usepackage{nicefrac}
\usepackage{mathtools, cuted}
\usepackage{lipsum}
\usepackage{float}
\usepackage{afterpage}
\usepackage{placeins}
\usepackage{subcaption}
\usepackage{tabularx}
\usepackage{balance}
\usepackage{textcomp}%
\usepackage{booktabs}
\usepackage{multirow}
\usepackage{accents}
\usepackage{svg}
\usepackage{url}
\usepackage{tikz}
\usepackage[american, straight voltages]{circuitikz}

\ctikzset{
  bipoles/thickness=1.1,
  nodes width=0.08
}

\usetikzlibrary{patterns,shapes,arrows}
\usetikzlibrary{positioning}
\allowdisplaybreaks
\ctikzset{
  bipoles/thickness=1.2,
  nodes width=0.08
}

\tikzstyle{decision} = [diamond, draw, fill=blue!20,
    text width=7em, text badly centered, inner sep=0pt]
\tikzstyle{decision1} = [diamond, draw, fill=blue!20,
    text width=7em, text badly centered, inner sep=0pt, node distance = 2.5cm]
\tikzstyle{block} = [rectangle, draw, fill=blue!20,
    text width=13em, text centered, rounded corners, minimum height=3em]
\tikzstyle{block2} = [rectangle, draw, fill=blue!20,
    text width=5em, text centered, rounded corners, minimum height=3em, node distance = 4cm]
\tikzstyle{block3} = [rectangle, draw, fill=red!20,
    text width=5em, text centered, rounded corners, minimum height=3em, node distance = 4cm]
\tikzstyle{line} = [draw, very thick, color=black!50, -latex']
\tikzstyle{cloud} = [draw, ellipse,fill=red!20, node distance=2.5cm, minimum height=2em]

\newcommand{\mb}{\mathbf}

\usepackage[nolist]{acronym}
\usepackage{comment}
\begin{acronym}
\acro{OCP}{Optimal Control Problem}
\acro{PMP}{Pontryagin's Minimum Principle}
\acro{MP}{Mathematical Programming}
\acro{QP}{Quadratic Program}
\acro{mpQP}{Multi-Parametric Quadratic Programming}
\acro{SQP}{Sequential Quadratic Programming}
\acro{LP}{Linear Programming}
\acro{MPC}{Model Predictive Control}
\acro{SPD}{Symmetric Positive Definite}
\end{acronym}

\archivemonth{November}       % defaults to the month you compile; \archivemonth{} hides it
\archiveyear{2026}            % defaults to the year you compile

\allowdisplaybreaks

\title{ Model Predictive Control on Analogue Hardware: \\ A Dual-flow Approach}

\author{Nikilesh Ramesh, Ross Drummond, Pablo R Baldivieso-Monasterios, Morgan Jones, \\ Yuanbo Nie% <-this % stops a space
\thanks{This work was supported by an EPSRC Doctoral Training Partnerships EP/W524360/1.}% <-this % stops a space
\thanks{The authors are with the School of Electrical and Electronic Engineering, University of Sheffield, S1~3JD Sheffield, U.K. (Emails:~{\tt{\{nramesh2,\,ross.drummond,\,p.baldivieso, morgan.jones},\,y.nie\}@sheffield.ac.uk}.)}%<-this % stops a space
}

\begin{document}

\maketitle

\begin{abstract}
Current Model Predictive Control (MPC) implementations are predominantly digital and, despite offering high precision and flexibility, often face challenges with ultra-low latency and hardware certification for safety-critical applications. We present an approach for solving the \ac{QP} of MPC using a continuous-time neural ordinary differential equation in the dual space. We obtain the solution to these QPs by mapping the differential equation's equilibrium point back to the primal space via a nonlinear transformation. The underlying dual-flow dynamics operate natively on analogue hardware. We validate this approach by demonstrating the hardware implementation of linear MPC on ``The Analog Thing" analogue computer. Finally, we extend the framework to nonlinear MPC with successive linearisation, broadening the applicability beyond linear systems.

\end{abstract}

\section{Introduction}
Model Predictive Control (MPC) has become a cornerstone of modern control because of its ability to handle constraints explicitly and optimise dynamic performance across diverse applications. The optimisation problem in MPC, for a given initial state $\tilde{x}$ and references $(y_\text{ref},u_\text{ref})$, is typically formulated as: 

\begin{align} 
   & \mathbb{P}(\tilde{x},y_\text{ref},u_\text{ref})\colon\min_{y_k,\,u_k}
    \;
      \tfrac{1}{2}\bigl\|y_{\text{ref}} - {y}_k\bigr\|_{{Q}}^2
    + \tfrac{1}{2}\bigl\| u_{\text{ref}} - {u}_k\bigr\|_{{R}}^2 \nonumber\\ 
   & \text{s.t. }  x_{k+1} = f(x_k, u_k), \quad y_{k} = g(x_k),\quad x_0 = \tilde{x}, \nonumber\\
   & \qquad y_{\min} \leq y_k \leq y_{\max},
    \quad k = 0, \ldots, N-1, \nonumber \\
    & \qquad u_{\min}\leq u_k \leq u_{\max},
    \quad k = 0, \ldots, N-1, \label{eqn:lmpc}
\end{align}

\noindent where $y_k\in \mathbb{R}^{q}, x_k\in \mathbb{R}^{n}$ and $u_k\in \mathbb{R}^{m}$ are the outputs, states and controls, $N>0$ is the control horizon, $k\in \{0,\ldots,N-1\}$ represents a time step, and $Q$ and $R$ are weighting matrices. When $f(\cdot)$ and $g(\cdot)$ are affine functions, the optimisation problem is called \emph{Linear MPC} (LMPC); otherwise, it is \emph{Nonlinear MPC} (NMPC). Typically, the MPC problem is transcribed to a parametric \ac{QP}  in the LMPC case and to a nonlinear program (NLP) in the NMPC case, \emph{i.e.,} a Sequential Quadratic Program (SQP). 

\begin{figure}[t]
    \centering
    \includegraphics[width=0.8\linewidth]{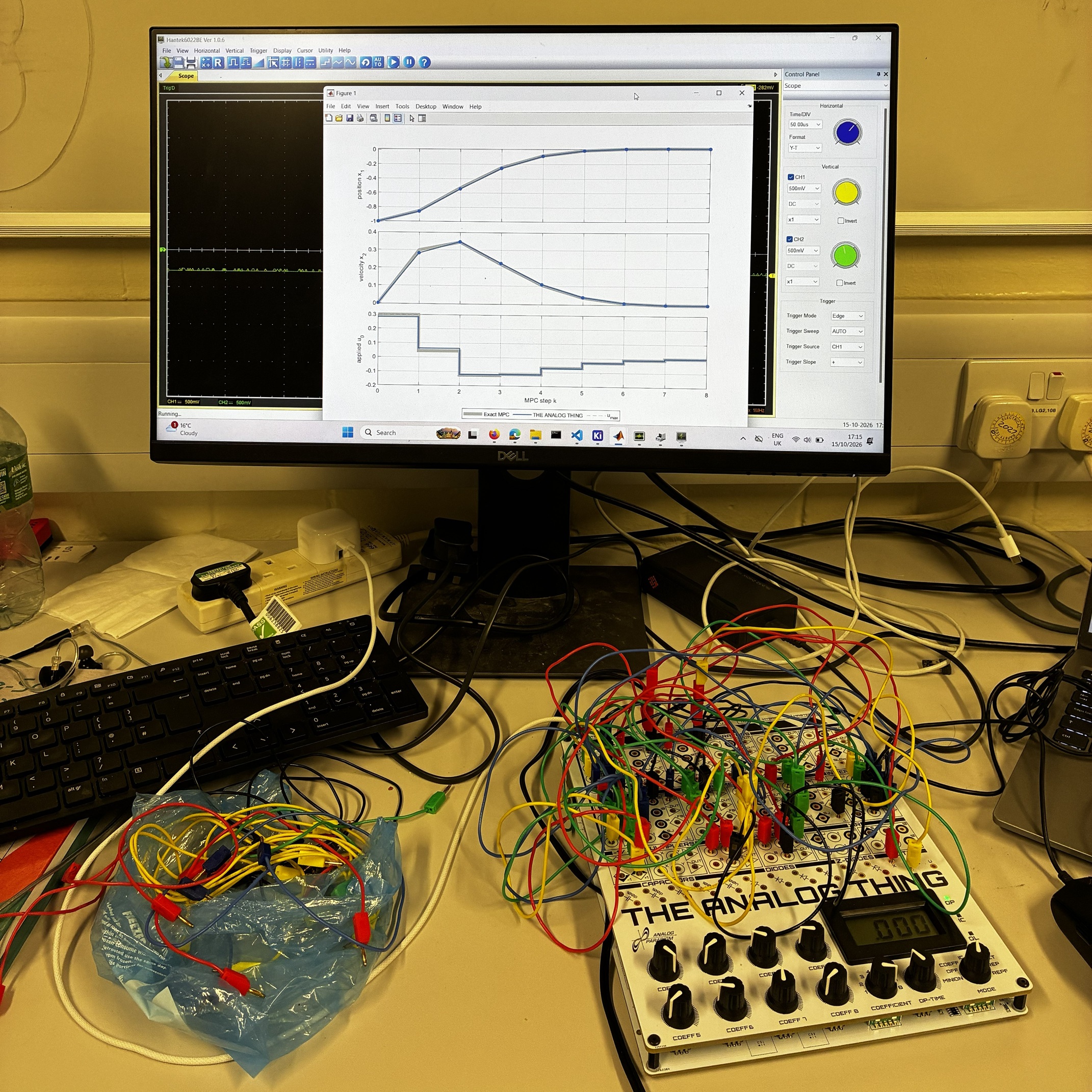}
    \caption{Picture of the ``\emph{The Analog Thing}" analogue computer setup used for the hardware demo of LMPC.}
    \label{fig:analogcomp}
\end{figure}

\subsection{Computation of the MPC solution}
There exists a vast literature on the algorithms for both linear and nonlinear MPC on digital computer systems \cite{Rawlings2017}. In this paper, we consider implementing MPC on analogue computers. Our implementation requires neither logic gates nor discrete bit representations to solve the QP \eqref{eqn:lmpc}. Instead, we compute entirely in continuous time via a neuro-dynamic reformulation of the QP subproblem; see Section~\ref{sec:algorithm} and the implementation in Figure \ref{fig:analogcomp}.

\begin{figure}[t]
    \centering
        \includegraphics[width=1\linewidth]{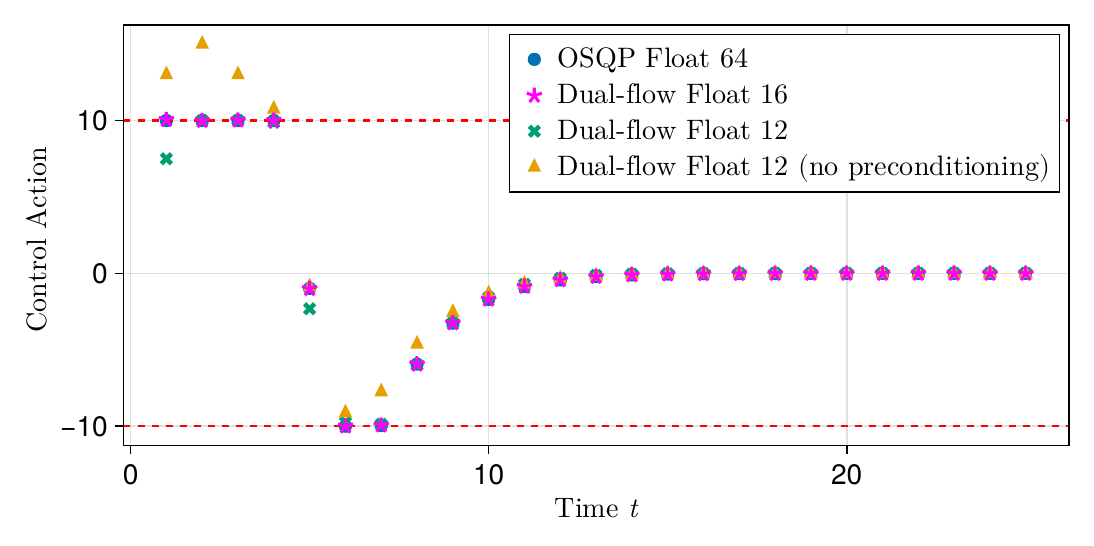}
    \caption{Comparison of control actions under varying precisions using dual-flow NDE for the example in \cite{drummond2024mappingforthmodelpredictive}. With 16-bit and preconditioned 12-bit align with OSQP (64-bit). Unpreconditioned 12-bit fails constraint satisfaction.}
    \label{fig:floatingpoint}
\end{figure}

Avoiding discrete computations can reduce computing time, making this approach suitable for real-time applications \cite{vichik2014solving}. Furthermore, the lack of a software layer may be desirable for safety-critical applications, as it avoids deadlocks or other logic bugs, and is robust against cyber-security attacks. However, algorithms built for analogue computation have issues, including: \emph{i)} the trade-off between numerical precision and the range of the  number representation (with this range defined by the magnitudes of the problem and the operational voltage range of the analogue device), \emph{ii)} tolerances in electric components introduce noise and biases,  and \emph{iii)} circuit complexity grows as a function of the problem size. We address these issues by modifying our implementation; these changes also yield practical benefits, such as enabling the use of lower-precision number representations, as shown in Figure \ref{fig:floatingpoint}.

\subsection{Related work on neuro-dynamic optimisation}

The proposed analogue QP solvers can be understood as an implementation of neuro-dynamic optimisation. This approach was pioneered by Hopfield in their analysis of a class of implicit networks that could ``\emph{store memory}" and solve optimisation problems \cite{Hopfield1985, HopfieldTank1986}. The continuous-time Hopfield network, generally defined as 
\begin{equation}\label{eqn:hopfield}
    \dot z(t) = -z(t) + W\Phi(z(t)) + b \quad \text{ s.t. } z(0) = z_0,
\end{equation}
has since yielded much attention in the field of mathematical programming. The earliest works in the neuro-dynamic optimisation literature were empirical studies that date back many decades, originating in nonlinear circuit theory and energy-based optimisation on analogue circuits \cite{Hopfield1985,Chua1984, HopfieldTank1986, Kennedy1988, FortiM.1995Ncfg}. Kennedy and Chua \cite{Kennedy1988} were among the first to establish a theory for solving NLPs on analogue circuits, mapping a penalty-based reformulation of the NLP to circuits to satisfy the Karush-Kuhn-Tucker (KKT) optimality conditions of the penalised problem. The primal states were represented as voltages, and the dual was estimated as current. The primal-space implementation required an ideal diode for exact representation, which results in primal infeasibility since ideal diodes do not exist.

Several works have extended the approach; for example, \cite{Ghasabi-OskoeiHasan2006Aesn, Tao2001} represented optimisation problems as Primal-Dual Neural Networks and solved saddle-point systems on circuits. More recently, primal and primal-dual circuits \cite{chaffey2025circuit, boyd2024optimization, vichik2014solving} have also been devised for optimisation. A potential issue is the need to solve for both primal and dual variables, which increases circuit complexity. 

We adapt dual-only networks \cite{ZHANG2002271, valmorbida2023quadratic}, the counterpart to the approach Kennedy and Chua \cite{Kennedy1988} originally considered.  Although this approach guarantees exact satisfaction of KKT conditions and avoids structural errors in the optimality conditions, it introduces distinct challenges: working in the dual space can lead to ill-conditioned weights and integration stiffness. Additionally, the system's dense weight matrix (from Hessian inverses) complicates the wiring effort and the handling of time-varying data required for SQP solvers. To mitigate these barriers for running MPC on analogue computers, we extend the dual network framework with additional enhancements, including a Hessian preconditioner and diagonal approximation.  

The remainder of this paper is organised as follows. Section \ref{sec:algorithm} introduces the dual-flow dynamics formulation for solving QP on analogue hardware. Section \ref{sec:example:LMPC} details the physical implementation of an LMPC benchmark on an off-the-shelf analogue computer. Finally, Section \ref{sec:nmpc} generalises the methodology to nonlinear systems using circuit simulations.

\section{A Dual-flow Algorithm  for analogue Compute} \label{sec:algorithm}

The parametric \ac{QP} for LMPC \eqref{eqn:lmpc} at a state $x_0$ is 
\begin{equation}  
 \min_{\mb{u}} \,\,  \frac{1}{2} \mb{u}^\top H \mb{u} +  x_0^\top F^\top  \mb{u} \quad  \text{s.t } \,   \, G\mb{u} \leq S_u \,x_0 +  w, \label{eqn:qp}
\end{equation}
where $\mb{u} = \{u_0,\ldots,u_{N-1}\}$. 
By introducing a change of variable $p = \mb{u} + H^{-1}Fx_0$, we can rewrite the problem as 
\begin{equation*}  
 \min_{p} \,\,  \frac{1}{2} p^\top H p  \quad  \text{s.t } \,   \, Gp \leq S \,x_0 +  w, \, \label{eqn:qp:rewrite}   
\end{equation*}
where $S = S_u + GH^{-1} F$. The KKT optimality conditions  associated  with this QP reduce to a Linear Complementarity Problem (LCP):
\begin{align*}
    z_i(S(i,\cdot)x_0 + w_i + G(i,\cdot)H^{-1} G^\top z) &= 0, \\
    z &\geq 0, \\
    Sx + w + G H^{-1} G^\top z &\geq 0,
\end{align*}
where $i = 1, 2,\dots m$ is row-wise iterator with $S(i,\cdot)$ representing the $i^\text{th}$ row of matrix $S$. In \cite{valmorbida2023quadratic}, this QP was shown to be equivalent to the following implicit relation: 
\begin{align*}
z &= -Sx_0 + (I - G H^{-1} G^\top) \text{ReLU}(z) - w, \\
    \mb{u} &= -H^{-1} F^\top x_0 - H^{-1} G^\top \text{ReLU}(z).
\end{align*}
This formulation can be understood as an implicit neural network \cite{drummond2024mappingforthmodelpredictive}. Further, assuming constraint qualifications hold and the objective is strongly convex, we will re-interpret the solution of the LCP as the equilibrium of the neuro-dynamic equation: 
\begin{subequations}\label{eqn:neurodynamic:qp}
\begin{equation}
    \dot z(\nu) =-z(\nu) +  W\,\mathrm{ReLU}(z(\nu)) +  b, \label{eqn:hopfield:qp}\\
\end{equation}
\noindent with $W = I - G H^{-1} G^\top$, and $b = -Sx_0 - w$. The variable $z$ represents the dual, and $\dot z(\nu)$ represents its flow over a virtual dimension $\nu$. Once the dynamical system reaches its equilibrium $z^*$, the QP solution $\mb{u}^*$ can be computed as 
\begin{equation}
      \mb{u}^* = -H^{-1} F^\top x_0 - H^{-1} G^\top \mathrm{ReLU}(z^*) \label{eqn:control:qp}.
\end{equation}
\end{subequations}

Hence, we have represented the solution of the parametric \ac{QP} \eqref{eqn:qp} as a neuro-dynamical flow in the dual space \eqref{eqn:hopfield:qp}. This system takes the form of the Hopfield network \eqref{eqn:hopfield}, but has explicitly defined weights and recovers the primal variables through the nonlinear map \eqref{eqn:control:qp}.  Hereinafter, we refer to \eqref{eqn:neurodynamic:qp} as the \emph{dual-flow}.

\subsection{Convergence analysis}

Convergence of the dual-flow of \eqref{eqn:hopfield:qp} to a steady-state solution directly affects the accuracy of the equilibrium point $z^*$ and, consequently, the control $u(\cdot)$ as well as the degree of constraint violation in the QP. These issues demonstrate a need to compute convergence certificates for \eqref{eqn:hopfield:qp}, with non-conservative methods including those based upon dynamic multipliers \cite{richardson2023strengthened, richardson2024strengthened, noori2025complete}. Here, the aim is to have a computationally efficient check so that the circuit's convergence can be rapidly evaluated. It is then proposed to apply results from euclidean contraction analysis of Hopfield networks \cite{Centorrino2023} for the convergence analysis. We observe the same assumptions and definitions of contraction as in \cite{FB-CTDS}.

\begin{lemma} \label{lemma:hopfield:contractivity}
    A continuous time Hopfield network \eqref{eqn:hopfield} with symmetric and non-singular weights $W$ is \emph{(i)} strongly contracting if $\lambda_\text{max} (W) < 1$ with rate $1 - \lambda_\text{max}(W)$ else if $\lambda_\text{max} (W) < 0$ with rate 1, or \emph{(ii)} weakly contracting if $\lambda_\text{max} (W) = 1$ (see Corollary 1 and Theorem 4 in \cite{Centorrino2023}).
\end{lemma}

\begin{lemma} \label{lemma:weak:contrax}
    Let $n$ and $m$ be number of primal variables and constraints associated with the parametric QP \eqref{eqn:qp}. Assuming \emph{Linearly Independent Constraint Qualification} holds the neuro-dynamic reformulation of QP \eqref{eqn:hopfield:qp} is weakly contracting by construction.  

    \emph{Proof:} Expand $\lambda_\text{max}(W) = 1 - \lambda_\text{min} (G H^{-1} G^\top)$, and since $H \succ 0$ implies that $H^{-1} \succ 0$. For $m > n$ then $\text{rank}(G) < m$ \emph{i.e.,} $G$ is rank deficient making $GH^{-1}G^\top \succeq 0$ singular. Hence,  $\lambda_\text{min} (GH^{-1}G^\top) = 0$ consequently $\lambda_\text{max}(W) = 1$ giving weak contraction from Lemma \ref{lemma:hopfield:contractivity}.
\end{lemma}
 
 The conditions of Lemmas \ref{lemma:hopfield:contractivity}-- \ref{lemma:weak:contrax} serve as  ahead-of-time checks for the convergence of the dual-flow. These checks are especially important when lower number representation are used since, as noted in Figure \ref{fig:floatingpoint}, the same dynamics of \eqref{eqn:hopfield:qp} when simulated in lower precision can fail to converge.

\paragraph*{Conditioning} A consequence of Lemma \ref{lemma:weak:contrax} is that the matrix $W$ is ill-conditioned, with the condition number denoted $\kappa(W) = \frac{1 - \lambda_\text{min}(GH^{-1}G^\top)}{1 - \lambda_\text{max}(GH^{-1}G^\top)}$ where $GH^{-1}G^\top \succeq 0$ and $\lambda_\text{max}(GH^{-1}G^\top) \neq 1$. Hence, the construction of $W$ in the dual-only framework can lead to stiff dynamics for   \eqref{eqn:hopfield:qp} which may converge slowly. This motivates the use of a preconditioner and constraint scaling for $G$  to enable useful solutions to be generated in  reduced arithmetic precision. We found that Jacobi preconditioning worked well in practice, especially in lower number representation, as shown in Figure \ref{fig:floatingpoint}.

\subsection{Analogue hardware implementation}
 
In the LTI case, the MPC algorithm can be completely represented with simple circuits built with Operational Amplifiers (op amps) and passive components.
Typically, the op amp takes two inputs ($V^+$ and $V^-$) that produce inverted and non-inverted outputs and  is powered by bi-polar supply voltages ($V_s^+$ and $V_s^-$). With additional passive components, the op amp can perform simple operations such as integration, inversion, and comparison. To implement the dual-flow of \eqref{eqn:neurodynamic:qp}, we require four main operators: \emph{(i)} summers, \emph{(ii)} inverters, \emph{(iii)} integrators and \emph{(iv)} ReLU. We will only give a brief note on the implementation of the ReLU operator here, as the first three are standard analogue circuits accessible in reference electronics  textbooks \cite{fiore2018operational}.

\subsubsection{ReLU operator} 
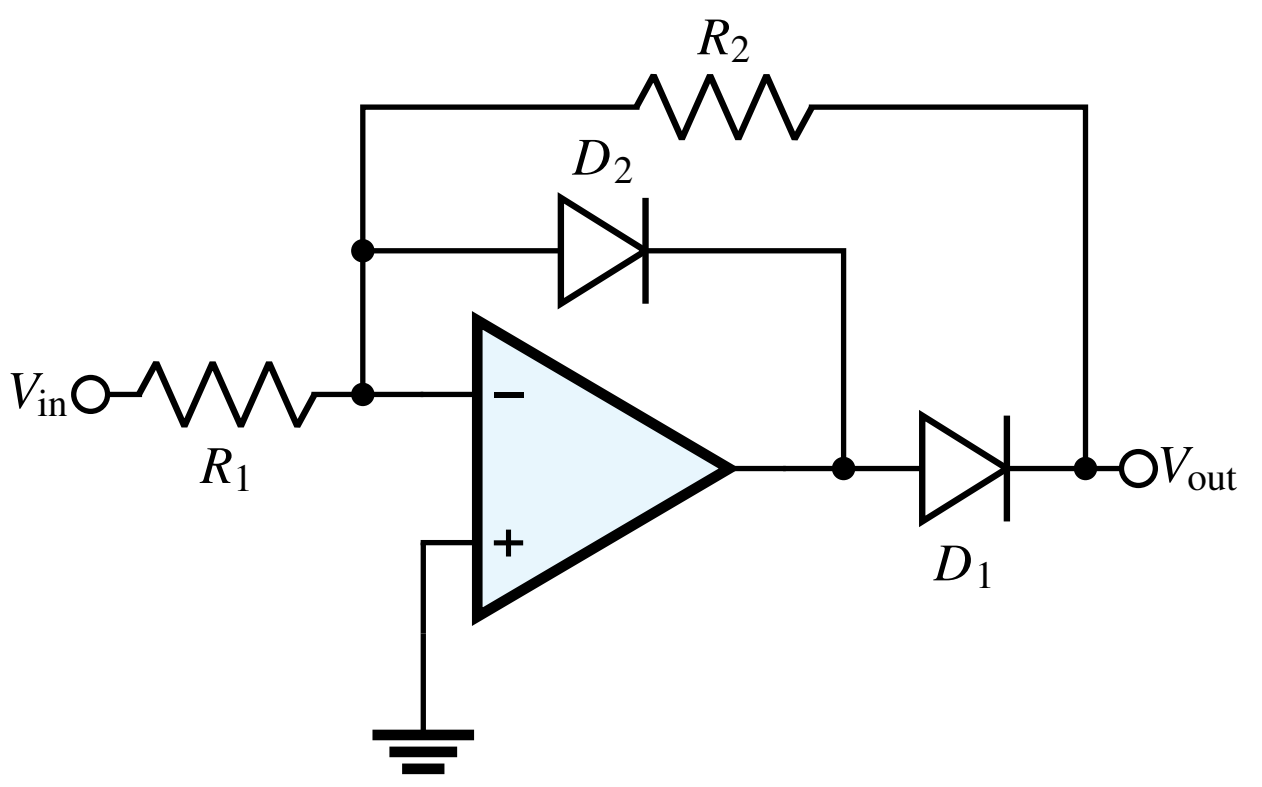
\begin{figure}[t]
    \centering
\begin{tikzpicture}[line width=1.0pt, scale=1, transform shape]
  % Op-amp node with light blue fill
  \node[op amp, fill=cyan!8] (opamp) at (0,0) {};

  % Non-inverting (+) input connected to Ground node with custom line width (0.6pt)
  \draw (opamp.+) -- ++(0,-0.6) node[ground, scale=1.6, anchor=center, line width=1pt] {};
  
  % Inverting (-) input node
  \draw (opamp.-) -- ++(-0.4,0) coordinate (inv_node);
  \fill (inv_node) circle (2.2pt);

  % Input resistor R1 and terminal Vin
  \draw (inv_node) to[R, l=$R_1$, -o] ++(-1.8,0) node[left] {$V_{\text{in}}$};

  % Branch node above inverting input
  \draw (inv_node) -- ++(0,0.95) coordinate (d2_in);
  \fill (d2_in) circle (2.2pt);

  % Op-amp output node
  \draw (opamp.out) -- ++(0.4,0) coordinate (out_node);
  \fill (out_node) circle (2.2pt);

  % Diode D2 branch
  \draw (d2_in) to[D, l=$D_2$] (d2_in -| out_node) -- (out_node);

  % Diode D1 branch and Vout node
  \draw (out_node) to[D, l_=$D_1$] ++(1.6,0) coordinate (vout_node);
  \fill (vout_node) circle (2.2pt);

  % Output terminal Vout
  \draw (vout_node) to[short, -o] ++(0.35,0) node[right] {$V_{\text{out}}$};

  % Feedback resistor R2 branch
  \draw (d2_in) -- ++(0,0.95) coordinate (r2_in)
        to[R, l=$R_2$] (r2_in -| vout_node) -- (vout_node);

\end{tikzpicture}
    \caption{{Circuit diagram of a precision half-wave rectifier representing the ReLU function used in \eqref{eqn:neurodynamic:qp}.}}
    \label{fig:relu}
\end{figure}

The ReLU function can be implemented with a precision half-wave rectifier as shown in Figure \ref{fig:relu}. When $V_\text{in} > 0 $ no current passes through the resistors otherwise the circuit is conducting. Hence the transfer characteristics of this circuit is $V_\text{out} = -\frac{R_2}{R_1} V_\text{in}$ when $V_\text{in} < 0 $ and $V_\text{out} = 0$ otherwise i.e. when $R_2 = R_1$ ($10$ k$\Omega$ chosen here) the circuit behaves as $V_\text{out} = -\text{ReLU}(V_\text{in})$. For high accuracy implementations, factors such as slew-rates and offsets affect the design. Further guidance is in~\cite{fiore2018operational}.

\subsection{Mapping dual-flow to op amp circuits} Solving the vectorized equation \eqref{eqn:hopfield:qp} to steady-state gives the solution to the QP,  with this system realisible by analogue circuit components. A summing integrator has the form $\tau \, \dot V = -\sum_j w_j V_j$ where $\tau$ is the time constant and the weight parameters are captured through resistor values. For mapping equation \eqref{eqn:hopfield:qp} to this form, isolate the $i$-th element of the dual-flow dynamics and expand the brackets:
\begin{equation*}
\begin{split}
    \tau \dot z_i(\nu) =-z_i(\nu) &+ W(i,\cdot)^\top \, \mathrm{ReLU}(z(\nu)) +b_i,
\end{split}
\end{equation*}
where $b_i = -  S(i,\cdot)^\top \, x_0 -  w_i.$ The  time derivative $ \dot z_i(\nu)$ can now be expressed using summer blocks. This implementation means that, for a problem with $m$ constraints, we will need $m$ summing integrators in parallel. The negative weight values are implemented by flipping the signs of the signal using inverter blocks. Note that a voltage reference is required to add constants to the summing integrator. Also, the dimension scaling of the problem is crucial since all the variables and constants, be they primal, dual, constraints or measured states, must all be representable within the available voltage range. The circuit implementation of poorly scaled problems can be more susceptible to noise induced errors. All the examples in this paper were manually scaled and tested before circuit implementation. Hardware dependent automatic scaling is theoretically possible since the duals of a well-posed QP can be bounded, but that's beyond the scope of this work.

\paragraph*{Hardware Scaling} In the LTI case, the QP's weights are fixed. Therefore, the circuit for the dual-flow algorithm can be represented with fixed resistor values.  However, if the system parameters change in time, the algorithm's circuit components must  be updated with measurements or parameter changes. Consider the case when Jacobian linearisation converts a nonlinear system into  a linear-parameter varying system, causing both $H$ and $c$ to change upon each iteration. To implement MPC on this system, an update map for the circuit components has to be built,  which could inflate the complexity. This added complexity is especially problematic since the inverse $H^{-1}$ increases the \emph{fill-in}, resulting in a dense $W$ matrix. To reduce the number of components in the circuit implementation,  Gerschgorin's circle theorem is used. This theorem has been widely used in large-scale regression type problems \cite{Bonte2018} but, to the best of our knowledge, has not been used in the context of MPC problems. The method yields a \emph{strictly diagonal} approximation $D$ of the Hessian:
\[
    D = \text{diag} (h_1, h_2, \dots h_i) \text{  for } i = 1,2,\dots n, \label{eq:Hinv}
\]
where $h_i = h_{ii} + \sum_{j \neq i}^n |h_{ij}|$ is the row-sum of absolute entries. This approximation  reduces the inverse to a diagonal matrix, with the diagonal elements being the reciprocal of the elements of $D$ i.e. $(\frac{1}{h_1}, \frac{1}{h_2}, \dots, \frac{1}{h_i})$. Applying this approximation reduces the number of elements to represent the inverse of the hessian from $n \times n$ to $n$. 

\section{Example: Linear MPC on Analogue Hardware} \label{sec:example:LMPC}
A toy problem with double integrator dynamics is set up to test the analogue setup using `\emph{The Analog Thing}' (\texttt{THAT}) by Anabrid, as seen in Figure \ref{fig:analogcomp}. The MPC is defined as:
\begin{align*}
    \min_u \sum_{k=0}^{N-1} x_k^\top Q x_k &+ u_k^\top R u_k \\
    \text{such that } x_{k+1} &= Ax_k + Bu_k,
\end{align*}

\noindent where $Q = \begin{bmatrix}
        1 & 0\\
        0 & 1
    \end{bmatrix}$, $B = \begin{bmatrix}
    \frac{1}{2}\Delta t^2 \\ \Delta t
\end{bmatrix}$, 
$A = \begin{bmatrix}
    1 & \Delta t \\
    0 & 1
\end{bmatrix}$ and $R = 1$ with $N = 2$ and $\Delta t = 1$ is the time step. A short horizon was chosen due to the hardware constraints as no more than a horizon of two would fit on this educational grade analogue computer. As the hardware is quite limited, the primary goal of this example is to demonstrate the algorithm working on real analogue device as a proof-of-concept.

\label{para:hardware} The analogue computer \texttt{THAT} is a general purpose device built for teaching. {As such, it lacks some of the components needed to fully implement the dual-flow algorithm of \eqref{eqn:neurodynamic:qp}}. For example, it is missing rectifiers. To mitigate this hardware constraint,  the \texttt{THAT}'s {connections} were adjusted to make use of the components available and solve the QPs. In particular, a comparator and multipliers were used to emulate rectifier behaviour. The \texttt{THAT} hardware uses knobs on potentiometers to set coefficient values for the weights and inputs. These knobs had to be manually tuned at each time step $k$ to set the measurement $x_k$ of the plant, with the steady-state voltage measurements then read off using a PC oscilloscope. The MPC loop was then ran in a MATLAB script, with the process  then repeated for the next time step $k$. Manually updating the potentiometer knobs on each time step is  a limitation of the current hardware implementation. However, if this hardware were to be deployed in the field, these values would be set automatically, either via digital switches or  from directly reading sensor values-- as done in the next example in Section \ref{sec:nmpc:example}.

\begin{figure}
    \centering
    \includegraphics[width=1\linewidth]{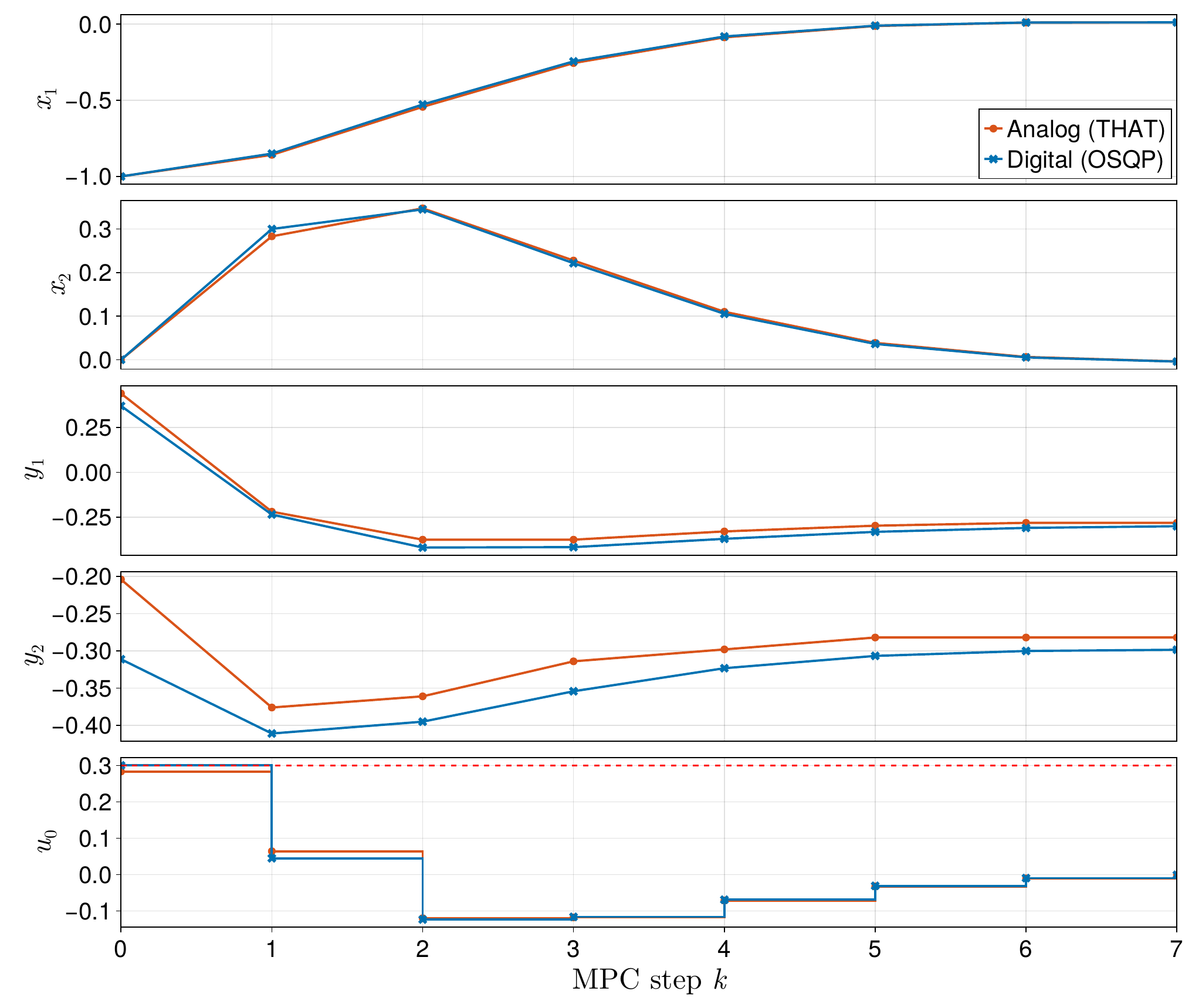}
    \caption{Results from hardware tests on \texttt{THAT} computer compared against trajectories obtained from \texttt{OSQP} \cite{osqp} on digital computer.}
    \label{fig:analoghardwareresult}
\end{figure}

The results from the analogue hardware implementations shown in Figure \ref{fig:analoghardwareresult} behave as expected; constraints are satisfied and the analogue primal trajectory remains  close to the trajectory computed by a digital computer. A larger gap  between the two trajectories is observed  in the dual plot, but this gap gets smaller after a few MPC steps. Tracing the sources of this gap from the  \texttt{THAT} hardware implementation is quite difficult, but a few ideas are proposed: 
\begin{itemize}
    \item The tuning knobs of the  potentiometers can only represent three digit significant figures in the interval [0,1]. They were also sensitive to disturbances (e.g. from human nudges), which made the manual tuning process error prone. 
    \item Since the hardware is meant for education demonstrations, the time constants are slow and passive components are of lower tolerances. For example, a reference $-1\;V$ signal with the potentiometer gain set to $0.3$ read $-0.29\;V$ instead of $-0.3\;V$. 
\end{itemize}

\section{Extension to Nonlinear MPC} \label{sec:nmpc}

Consider the case of an MPC problem with a nonlinear model. The presented QP algorithm from \eqref{eqn:neurodynamic:qp} for LMPC can be extended to this more complex setting by solving the optimisation problem using SQP. With the SQP approach, a sequence  of $\ell = 1,\,\dots,L$  QPs are solved at each MPC time step:
\begin{equation*}
 \min_{\Delta \mb{u}_{\ell}}  \frac{1}{2} \Delta  {\mb{u}_{\ell}}^\top H_\ell \Delta  \mb{u}_{\ell} +  {c_\ell}^\top \Delta  \mb{u}_{\ell}, \, \text{ s.t } \, G_\ell \,  \mb{u}_{\ell+1} \leq h_\ell, \, \label{eqn:sqp}    
\end{equation*}
with the iterates updating by $\mb{u}_{\ell+1} = \mb{u}_ {\ell} + \alpha \Delta \mb{u}_{\ell}$ for some $\alpha \in (0,1]$, and the QP data $H_\ell, c_\ell, G_\ell, h_\ell$  re-evaluated at each iterate. The iteration is initialised with the solution from the previous sampling instant, shifted by one step, and the first element $u_0$ of $\mb{u}_L$ is applied to the system.

In conventional SQP, line-search methods are typically used for globalisation \cite{nocedal1999numerical}. Using line searches here, however, is challenging because they must evaluate trial points and compare them against a merit function for acceptance at each iteration. This is undesirable for analogue implementations since it increases circuit complexity. Instead, we use ideas from \emph{Real-time Iteration} (RTI) \cite{Diehl2005}. Specifically, we do not implement a line search but apply one QP solve per MPC step. The validity of this approach is due to the availability of ``\emph{good}" warm-started solutions for feedback control problems, since the solutions update at such high frequencies that, typically, the active set changes slowly.

\paragraph*{Limitations:} In this work, we limit our attention to the class of nonlinear models that, when Jacobian linearisation is applied, yield an LTI system and have no output, state, or time-varying constraints. This is primarily because explicit time-dependence would require introducing a new variable for time, which would grow unbounded. Active set changes also allow for structural changes in the data matrices, but implementing this on the analogue computer would require online circuit reconfiguration, thus increasing hardware complexity. Apart from circuit complexity, active set changes in SQP may also cause recursive feasibility issues in ill-posed problems. Various fixes have been proposed to alleviate these feasibility issues but implementing them on analogue hardware remains an open problem. We leave the handling of these active set changes for future work and only consider nonlinear problems where the set does not change. 

\subsection{Co-Simulation Design} \label{para:cosim} 
The plant and controller are co-simulated, with the plant dynamics simulated in the same circuit simulators and a connection that supplies measurements to the controller. Nonlinear MPC based on successive linearisation requires updating the optimisation problem data matrices around the current linearisation point at every MPC step using plant measurements. In this context, the Hessian and Jacobian can be constructed from measurement-dependent matrix polynomial maps. Consider the Jacobian matrix $\bar M$  defined by 
\begin{equation}
  \label{eqn:matrixpoly}
     \bar M = \begin{bmatrix}
     C \bar B & 0 & \dots  & 0 \\
     C \bar A \bar B & C \bar B & \dots & 0 \\
    \vdots & & & \\
    C \bar A^N \bar B & C \bar A^{N-1} \bar B & \dots & C \bar B
 \end{bmatrix}
\end{equation}
where $\bar A$ and $\bar B$ are the linearised discrete system matrices around the measurements at the current step. For example, a degree 2 matrix polynomial may be constructed as $\bar M = M_0 + M_1 \bar x_0 + M_2 {\bar x_0}^2$, with $\bar x_0$ the state measurement $\bar M$ depends on, making the construction problem specific. Section~\ref{sec:nmpc:example} offers more details where the plant and controller circuits are co-simulated using their netlists and the \texttt{NGSPICE} simulator \cite{ngspice}.

\subsection{Nonlinear MPC Example} \label{sec:nmpc:example}

Consider a reference tracking NMPC problem with a time-varying trajectory and a nonlinear bicycle model,  defined by: 
\begin{align*}
    \min_{y(\cdot),u(\cdot)} \int_{t_0}^{t_f}(y(t) &- \bar y(t))^\top Q (y(t) - \bar y(t)) + u(t)^\top R \, u(t) \, dt \\
    \text{subject to},&\text{ for all }  t\in[t_0,t_f]:\\
    \dot p_x(t) & = v(t) \cos(\psi(t) + \beta(t)), \\
    \dot p_y(t) & = v(t) \sin(\psi(t) + \beta(t)), \\
    \dot v(t) & = a(t), \\
    \dot \psi(t) &= \frac{v(t)}{\ell_r} \sin(\beta(t)),\\
    \beta(t) & = \arctan \bigg( \frac{\tan(\delta(t)) \, \ell_r }{\ell_f + \ell_r}\bigg), \\
    |a(t)| &\leq 2.5\text{, } |\delta(t)| \leq 0.8\text{, } |\dot \delta(t)| \leq 0.1.
\end{align*}
The states are $p_x(t), p_y(t), v(t)$, and $\psi(t)$, corresponding to lateral and longitudinal positions, velocity, and yaw angle. The output is $y(t) = [p_x(t), p_y(t)]^\top$. The control inputs are the acceleration and steering angle \emph{i.e.,} $u(t) = [a(t), \delta (t)]^\top$. The slide-slip angle $\beta(t)$ is calculated from $u_2(t)$ and the constant front and rear wheelbase lengths $\ell_f=1$ and $\ell_r=1.5$ unit meters. Additionally, a rate constraint is added to the steering.

Because the model uses trigonometric functions, we use small-angle approximations, $\sin(\theta) \approx 0$ and $\cos(\theta) \approx 1$, to simplify the analogue implementation. Keeping these nonlinear functions would require additional analogue function generators in the circuit, which would introduce more errors in the solution and increase circuit complexity. We construct discrete system matrices (with zero-order holds) from the linearised state-space equations to give the dynamics $x_{k+1} = \bar A x_k + \bar B u_k$, with
\[
 \bar A = \begin{bmatrix}
     1 & 0 & \Delta t & 0 \\
     0 & 1 & 0 & v_0 \Delta t \\
     0 & 0 & 1 & 0 \\
     0 & 0 & 0 & 1
 \end{bmatrix}, \bar B = \begin{bmatrix}
     \frac{\Delta t^2}{2} & 0\\
     0 &  \frac{(v_0^2 \Delta t^2 + 2\ell_r v_0 \Delta t)}{2(\ell_r + \ell_f)} \\
     \Delta t & 0 \\
     0 & v_0 \Delta t \frac{1}{\ell_r + \ell_f}
 \end{bmatrix}
\]
where $v_0$ is the velocity measurement at the current MPC step. In this case, \eqref{eqn:matrixpoly} depends only on the velocity measurement $v_0$ and its square. This matrix can then be written as a matrix polynomial $\bar M = M_0 + M_1 v_0 + M_2 {v_0}^2$. Since these individual matrices,  $M_0,\, M_1,$ and $ M_2$, are fixed, they can be represented by resistors. Analogue multipliers can then be used to compute ${v_0}^2$ from the measurement signal. Since the Hessian is $H = \bar M^\top Q \bar M + R$, we can use $\bar M_k$ to obtain another polynomial representation of the Hessian diagonal approximation. This Hessian approximation will be formed of higher powers of the signal, namely ${v_0}^2$, ${v_0}^3$ and ${v_0}^4$. This means more analogue multipliers are required, and inaccuracies can arise as we add more transformations to the same signal. Analogue sample-and-hold devices are used to store measurements $v_0$ from the plant (or alternatively can be thought of as from a velocity sensor), which are then used for the linearisation. The resulting sequence of QPs is then solved with the dual-flow algorithm of \eqref{eqn:hopfield:qp}. 

\begin{figure}[t]
    \centering
    \includegraphics[width=1\linewidth]{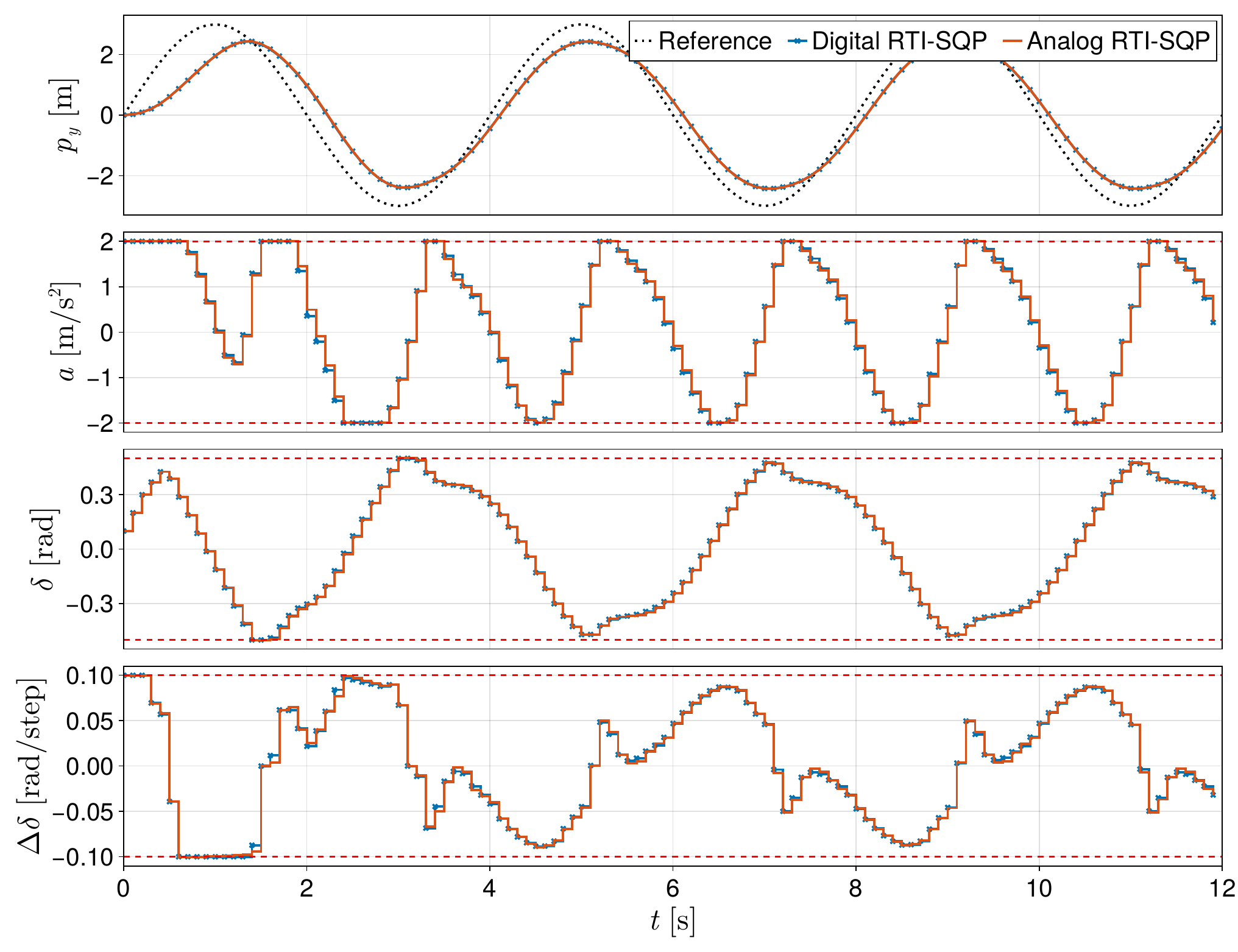}
    \caption{Comparison of a Digital RTI-SQP MPC trajectory against Analogue RTI-SQP implementation. All solutions use Gershgorin approximation for the Hessian, with Digital solved using \texttt{OSQP} and Analogue using dual-flow algorithm simulated in \texttt{NGSPICE}.}
    \label{fig:nmpc}
\end{figure}

Figure \ref{fig:nmpc} plots the nonlinear closed-loop trajectories from the analogue dual-flow controller. The figure shows trajectories with SQP running one iteration per MPC step (RTI) along with comparisons against digital versions of the same problem solved with the \texttt{OSQP} solver \cite{osqp}. The presented algorithm can be observed to perform well against the digital implementations, as seen in Figure \ref{fig:nmpc}, with the trajectories appearing identical.

\section{Conclusions and future work} \label{sec:concl}

We presented a dual-flow neuro-dynamic algorithm for analogue implementations of condensed model predictive control (MPC). The proposed approach is applicable to both linear time-invariant models with general affine constraints and nonlinear systems with input constraints. By representing the optimisation solution as an equilibrium point in the dual space, the primal solution can be recovered through a nonlinear transformation. This approach leverages contraction theory to guarantee convergence ahead of time, while allowing designers to tune the settling time via passive components--enabling higher control frequency without algorithmic redesign.

The presented approach was validated on a linear MPC problem implemented on an education-grade analogue computer, and a nonlinear MPC problem emulated with a circuit simulation. The results demonstrated the attractiveness of this approach despite the inherent noise and inaccuracies present in computing with analogue devices. Future work will focus on addressing limitations in handling active-set changes that alter constraint-matrix structure and scaling the implementation to higher-dimensional problems. We are also pursuing applications in safety-critical systems and extending the hardware validation to nonlinear controllers. 

\bibliography{references} 
\bibliographystyle{ieeetr}

\end{document}